\documentclass[12pt]{article}
\usepackage[intlimits]{amsmath}
\usepackage{amssymb,color}

\usepackage{psfrag,graphicx}
\makeatletter\def\input@path{{pics/}}\makeatother\graphicspath{{pics/}}

\DeclareMathOperator{\Tr}{Tr}

\newcommand{\acro}[1]{#1\@} 
\newcommand{\N}{{\cal N}}

\begin{document}
\baselineskip 18pt


\thispagestyle{empty}
\renewcommand{\thefootnote}{\fnsymbol{footnote}}

{\hfill \parbox{4cm}{
    MPP-2006-58 \\    SHEP-06-19 \\
}}

\bigskip\bigskip

\bigskip

\begin{center} \noindent \Large \bf
Heavy-Light Mesons from the AdS/CFT Correspondence
\end{center}

\bigskip\bigskip\bigskip

\centerline{ \normalsize \bf Johanna Erdmenger $^a$, Nick Evans
$^b$ and Johannes Gro\ss e $^a$ \footnote[1]{\noindent \tt
 jke@mppmu.mpg.de, evans@phys.soton.ac.uk, jgrosse@mppmu.mpg.de} }

\bigskip

\bigskip\bigskip

\centerline{ \it $^a$ Max-Planck-Institut f\"ur Physik
(Werner-Heisenberg-Institut)} \centerline{ \it F\"ohringer Ring 6,
80805 M\"unchen, Germany}
\bigskip

\bigskip\bigskip

\centerline{ \it $^b$ School of Physics \& Astronomy,
Southampton University} \centerline{\it  Southampton, S017 1BJ,
United Kingdom}
\bigskip

\bigskip\bigskip

\bigskip\bigskip

\renewcommand{\thefootnote}{\arabic{footnote}}

\centerline{\bf \small Abstract}
\medskip

{\small \noindent We propose a holographic description of
heavy-light mesons, i.e.~of mesons containing a light and a heavy
quark. In the semi-classical string limit, we look at the dynamics
of strings tied between two D7 branes. We consider this setup both
in an AdS background and in the non-supersymmetric Constable-Myers
geometry which induces chiral symmetry breaking. We compute the
meson masses in each case. Finally we discuss the relevance
of this result for phenomenological comparison to the physical b-quark sector.}

\newpage


\section{Introduction}

The AdS/CFT correspondence \cite{Mal,Gub,Wit} and its generalizations
offer the possibility to holographically describe the strong coupling
regime of \acro{QCD} using weakly coupled gravity.  There has been
steady progress towards such a description including string theory
descriptions of theories that break chiral symmetry \cite{BEEGK} --
\cite{Karch:2006bv}.  Most recently these ideas have been adapted to
produce phenomenological models of \acro{QCD} dynamics that describe
aspects of the meson and baryon spectrum at the 20\% level or better
\cite{Erlich:2005qh} -- \cite{Hambye:2005up}.

Within models based on AdS/CFT, quarks may be introduced by the
inclusion of D7 branes in the holographic space \cite{KarchKatz} --
\cite{Arean:2006pk}~\footnote{Related models involving other brane
  setups may be found in \cite{Nunez} -- \cite{Benvenuti:2005qb}.}.  A
small number of quarks, $\psi^i$, ($N_f \ll N_c$), may be described by
treating the D7 branes as probes \cite{KarchKatz, Mateos}.  In the
simplest case with the background geometry 
$\mathrm{AdS}_5 \times \mathrm{S}^5$, the
geometry around a stack of D3 branes, the resulting field theory is an
$\N=2$ gauge theory with fundamental hypermultiplets. On the
field theory side of the correspondence, the quarks are described by
strings stretching between the D3 and the D7 branes.  The quark mass
is proportional to the separation between the D3 brane stack and the
D7 probe. The holographic dual description of the quark bilinear
$\bar{\psi} \psi$ is given by open strings with both ends on the D7
brane probe in the $\mathrm{AdS}_5 \times \mathrm{S}^5$ background. Therefore
fluctuations of the D7 brane correspond to meson excitations. In
\cite{BEEGK}, a holographic description of chiral symmetry breaking
and light mesons was given by embedding a D7 brane probe into a
deformed non-supersymmetric supergravity background.

Here we push these ideas into a new arena and describe mesons made of
one light quark, $\psi_L$, which in \acro{QCD} would experience chiral
symmetry breaking dynamics, and one heavy quark, $\psi_H$, that would
not.  To begin with, we formulate our new approach by considering a
string theory description of the supersymmetric field theory dual to
$\mathrm{AdS}_5 \times \mathrm{S}^5$ with D7 probes wrapping $\mathrm{AdS}_5 \times \mathrm{S}^3$, which
is conformal when the quarks are massless.  To include a heavy quark
and a light quark, we must have {\it two} D7 branes with very
different separations from the central D3 brane stack. Heavy-light
mesons are described by the strings stretched between these two D7
branes. In the limit where one quark is very heavy, these strings
become long, and we may use a semi-classical description of their
dynamics. As the simplest ansatz, we just consider the motion of a
rigid string in directions transverse to the separation of the D7
branes, i.e.~in the $x$ directions of the field theory on the D3 brane
as well as in the radial holographic direction. In other words we
treat the D7-D7$'$ string as rigidly tied between the two D7 and
neither let it bend nor oscillate in the direction of its length. We
can then integrate over the string length in the string (Polyakov)
action.  In this way we obtain an effective point-particle-like action
for the string's motion. In the spirit of second quantization, we
convert the action to a field equation which we consider as the
holographic field theory description of the heavy-light mesons. From
the field-theory point of view, this equation is a generalization of
the Klein-Gordon equation~\footnote{In a somewhat different approach
  which uses spinning strings, multiflavour mesons were also
  considered in \cite{Paredes:2004is}. Further related work may be
  found in \cite{Pons:2004dk} -- \cite{Herzog:2006gh}.}.  Note that in
our semi-classical string theory formalism, vector and scalar meson
masses are inherently degenerate. On the field theory side, this ties
in well with the fact that spin effects are suppressed by $1/m_H$ with
$m_H$ the heavy quark mass \cite{Manohar}.

Note that the field theory describing the heavy-light mesons does not
exist in the same space-time as the fields holographically describing
light-light or heavy-heavy mesons, but in a space given by the
integrated average over the space between the two D7. For this reason
one needs a fully stringy picture of the D7 branes to describe these
states and they cannot be extracted from the simple five-dimensional
models found in \cite{Erlich:2005qh} -- \cite{Hambye:2005up}.

In this manner we present results for the heavy-light states in the
$\N=2$ supersymmetric gauge theory so far described.  However
this model does not include chiral symmetry breaking behaviour.
Therefore we move on to the authors' preferred string theory
description of chiral symmetry breaking \cite{BEEGK} in the
non-supersymmetric Constable-Myers background
\cite{CM,Gubser,add2,Gubser2}. This model consists of an AdS
background deformed by the presence of a non-zero dilaton. The
presence of the dilaton breaks the conformal and supersymmetries of
the string setup, leaving a non-supersymmetric strongly coupled
confining gauge background. The advantage of this background is that
in the UV, it returns to four-dimensional $\N=2$ theory (with the
field content of $\N=4$ plus fundamental matter), such that the
asymptotic behaviour of the brane embedding is well under control.

Quarks are again introduced by D7 brane probes in the deformed
geometry.  It is known from \cite{BEEGK} that the D7 branes are
repelled from the center of the geometry, triggering the formation of
a chiral symmetry breaking condensate. In order to obtain heavy-light
mesons and their masses, we use our new formalism developed in the
supersymmetric case.  This time however, for the deformed background,
we have to integrate numerically over the length of D7-D7$'$ string.

Phenomenological holographic models have taught us that such string
theory models work well at describing \acro{QCD}. We therefore allow
ourselves to be tempted into describing the \acro{QCD} b-quark mesons,
even though the b-quark is not infinitely massive and the gauge
background is not that of \acro{QCD}. The AdS description, which is fully valid
only at infinite 't Hooft coupling  $\lambda$, predicts that the 
light-light and heavy-heavy meson masses are suppressed relative to the 
constituent quark mass by $\sqrt{\lambda}$. 
There is relatively little evidence for this behaviour in real
\acro{QCD}.
However, in this limit we find that 
the mass of the lowest of the 
heavy-light states coincides with the  heavy quark mass.  We use the 
$\rho, \Upsilon$ and $B$ meson masses to fix $\Lambda_{QCD}$, the
heavy quark mass $m_H$
and $\lambda$ in our model. The $\rho$ and $\Upsilon$ mesons are
light-light and heavy-heavy vector mesons, respectively. 
This procedure generates the result $\lambda \sim 5$ and
allows for a prediction of the
$B^*$ meson mass, which however yields a result which is
$20\%$ too high. The observed
$B-B^*$ splitting can only be generated with a much larger value of $\lambda$ 
which makes the $\rho, \Upsilon$ too light. The model does provide
a rough caricature of the \acro{QCD} states though.

\section{Mesons in AdS}

\subsection{Probe D7 Branes in AdS}

\begin{figure}[t]
\begin{center}
\begin{tabular}{cc}
  \includegraphics[height=4cm]{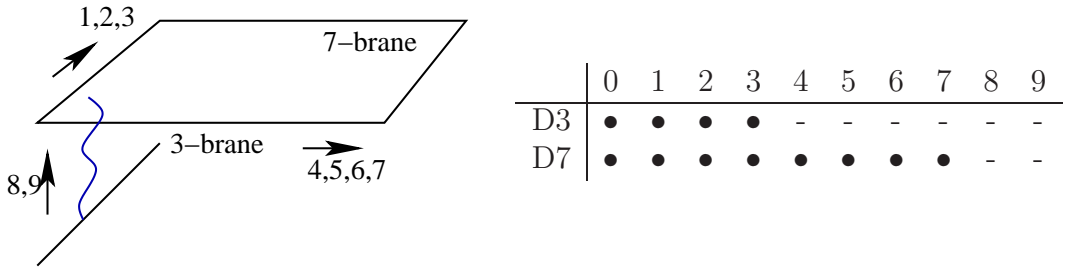}
\vspace{-3cm}& \\ & 
  \begin{tabular}{c|cccccccccc} 
  & 0 & 1 & 2 & 3 & 4 & 5 & 6 & 7 & 8 & 9 \\ \hline  
  D3 & $\bullet$ &$\bullet$ &$\bullet$ &$\bullet$ & -& -& -& -& -& -\\
  D7 & $\bullet$ &$\bullet$ &$\bullet$ &$\bullet$ & $\bullet$
     & $\bullet$& $\bullet$ &$\bullet$ & -& -
  \end{tabular} 
\end{tabular}
\vspace{1.5cm}
\caption{The basic geometry of the D3-D7 system under
  consideration.\label{fig1}}
\end{center}
\end{figure}

Quark fields may be introduced into the $\N=4$ gauge theory described
by the AdS/CFT correspondence by including probe D7 branes.  The
resulting theory is an $\N=2$ theory investigated previously in
\cite{KarchKatz,Mateos}. Neglecting the back reaction of the D7 in the
probe limit corresponds to considering the quenched approximation on
the field theory side.  As displayed in figure \ref{fig1}, strings which
stretch between the D3 and D7 branes are in the fundamental
representation of the $\mathrm{SU}(N)$ gauge theory on the D3. The
length of the minimum length string between the two branes determines
the mass of the quark field.

The gravity dual of the D3 branes is $\mathrm{AdS}_5\times \mathrm{S}^5$, in which the D7 brane
probe wraps $\mathrm{AdS}_5 \times \mathrm{S}^3$. This configuration minimizes the world-volume
action of the D7 probe.

The AdS metric is usually written as 
\begin{equation}   \label{AdSmetric} 
  ds^2 = \frac{w^2 }{R^2} \eta_{\alpha \beta} dx^\alpha dx^\beta 
       + \frac{R^2}{ w^2} ( d \rho^2 + \rho^2 d\Omega_3^2 + dw_5^2 + dw_6^2)\,, \hspace{1cm} 
  w^2= \rho^2 + w_5^2 + w_6^2\,,
\end{equation} 
where $\Omega_3$ corresponds to a three-sphere and $R^4=4 \pi g_s N
\alpha'{}^{2}$. Note that scale transformations in the field theory,
which define the mass dimension of operators (for example if we scale
$x \rightarrow e^h x$ then a scalar field of dimension one scales as
$\phi \rightarrow e^{-h} \phi$), are mapped to a symmetry of the
metric with the radial direction, $w$, scaling as an energy scale.


We place the D7 brane so that its world-volume coordinates are
$\xi^a=x^\alpha, \rho, \Omega_3$. Strings with both ends on the D7
brane generate the DBI action for the brane
\begin{equation} \label{DBI} 
  S_{D7} = -T_7\int d^8\xi~e^\phi\left[-\det({\bf P}[g_{ab}])\right]^{\frac{1}{2}} \,,
\end{equation}
where the pull-back of the metric ${\bf P}[g_{ab}]$ is given by
\begin{equation}
  {\bf P}[g_{ab}] = g_{MN} \frac{dx^M}{d \xi^a} \frac{dx^N}{d \xi^b} \, .
\end{equation}

In ${\mathrm AdS}_5 \times {\mathrm S}^5$ this gives 
\begin{equation}
  S_{D7} = - T_7 \int d^8\xi\, \epsilon_3 \, \rho^3 
    \sqrt{1 + \frac{ R^2 g^{ab} }{ \rho^2 + w_5^2 + w_6^2} 
                  (\partial_a w_5\partial_b w_5 + \partial_a w_6\partial_b w_6)}\, , 
\end{equation}
where $g_{ab}$ is the induced metric on the D7 and $\epsilon_3$ is the
determinant factor from the 3-sphere. Moreover $T_7=(2 \pi^7 g_s
\alpha'{}^4)^{-1}$.

The regular D7 brane embedding solutions are just $w_5^2+w_6^2=d^2$,
i.e.~the brane lies at a constant radius in the $w_5-w_6$ plane with
$m = d /(2 \pi \alpha') $ the quark mass. $d$ is the length
  of the shortest D3-D7 string. 

Fluctuations of the D7 brane (which asymptotically fall off as
$1/\rho^2$) are dual to meson fields made of the fermionic quark and
anti-quark. If we work in the particular choice of background
embedding $w_5=0$, $w_6=d$ and parametrize fluctuations as
\begin{equation} 
   w_6 + i w_5 = d + \delta(\rho) e^{ik\cdot x}, \hspace{1cm} M^2=-k^2 \, , 
\end{equation}
then the linearized equation of motion for $\delta$ is
\begin{gather}
  \partial_\rho^2 \delta(   { \varrho} ) 
    + \frac{3}{\varrho} \partial_\varrho \delta(\varrho) 
    + \frac{\bar M^2}{(\varrho^2 + 1)^2} \delta(\varrho) =0 \, , \quad 
  \varrho \equiv  \frac{\rho}{d} \, , \quad 
  \bar M^2 \equiv - \frac{k^2R^4}{ d^2} \,.
\end{gather}
The solution is given in terms of hypergeometric functions
\cite{Mateos}, 
\begin{equation} 
  \delta(\rho) = \frac{1}{(\rho^2 + d^2)^{n+1}} F(-n-1,-n;2,-\rho^2/d^2) \, . 
\end{equation}  
The mass spectrum of the degenerate scalar and pseudo-scalar states is
then given by
\begin{equation} \label{llmass} 
  M = \frac{2 d}{R^2} \sqrt{(n+1)(n+2)}, \hspace{1cm} n=0,1,2,\dots
\end{equation}

Note that the masses of states scale as $d/R^2 = m (2 \pi
\alpha')/R^2$, which is linear in the quark mass, independent of
$\alpha'$ and scales as one over the square root of the 't Hooft
coupling (since $R^4/(2\pi\alpha')^2 =g_s N/\pi$). In the limit
$d=0$ the theory becomes conformal and there is not a discrete
spectrum.

If we have two D7 branes embedded at $w_6=d,D$ respectively then there
are meson states of the form ``light-light" or ``heavy-heavy" with the
ratio of masses of the lightest two mesons just $d/D$.

\begin{figure}[t]
\begin{center}
  \includegraphics[height=4.5cm]{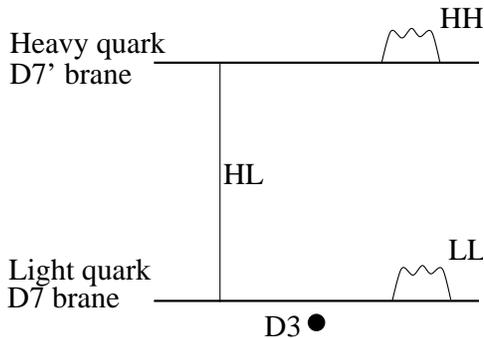}
  \caption{The brane configuration including both a heavy and a light quark.
  The 77 and $7'7'$ strings are holographic to light-light and heavy-heavy
  mesons respectively. Heavy-light mesons are holographically described by
  strings between the two D7 branes -- we work in the semi-classical limit
  where those strings are stretched tight.}\label{f:stretch}
\end{center}
\end{figure}

\subsection{Heavy-Light Mesons in AdS\label{hlm}}

We describe mesons made of one heavy and one light quark by strings
stretched between the two D7 branes. In the limit $D \gg d$, with $D$,
$d$ the distances of the heavy and the light quark brane from the
D3's, these strings are very long such that it is appropriate to
consider them in the semi-classical limit (see fig.~\ref{f:stretch}).

All of the fields that make up the (non-abelian) DBI action of the
D7 branes come from dimensional reduction of ten-dimensional gauge
fields. These gauge fields originate from the lightest quantum
state of the open strings connecting the D7 branes. Strings with
both ends on the inner D7 describe the light quark operators
${\bar \psi}_{L} \psi_L$, ${\bar \psi}_{L} \gamma_5 \psi_L$,
${\bar \psi}_{L} \gamma^\mu \psi_L$. Similarly the strings with
both ends on the outer D7 describe the heavy quark states. In the
scenario proposed here, the strings stretched between the two D7
branes describe the heavy-light modes ${\bar \psi}_{H} \psi_L$,
${\bar \psi}_{H} \gamma_5 \psi_L$ and ${\bar \psi}_{H} \gamma^\mu
\psi_L$. Strings of this type without surface oscillations are the
tachyons projected out by the GSO projection. In the
semi-classical limit we employ, the string mass is assumed to be
dominated by the string length and the subtleties of quantum
corrections and surface fluctuations are ignored. The unexcited
string stretched between the two D7 branes which we study is
therefore an approximation to the scalar, pseudo-scalar and vector
heavy-light states which  by assumption are degenerate.

We
use the gauge-fixed Polyakov string action
\begin{equation} 
   S_P = - \frac{T}{2} \int d\sigma\, d \tau \; G_{\mu \nu} 
           ( - \dot{X}^\mu \dot{X}^\nu + X'{} ^{\mu} X'{} ^{\nu}) \, , \qquad 
  T=\frac{1}{2\pi\alpha'}\,,
\end{equation}
so we must also impose the constraint equations
\begin{equation} 
  G_{\mu \nu} \dot{X}^\mu X'{}^\nu =0\,, \hspace{1cm} 
  G_{\mu\nu}(\dot{X}^\mu \dot{X}^\nu + X'{}^\mu X'{}^\nu)=0  \, . 
\end{equation}

Our string will lie between the two D7 branes so that
$\sigma=w_6$. We will allow the string to move in the world volume
of the D7 brane. We thus integrate over the $w_6$ direction in
order to generate an effective point particle action. With $X^\mu
= \{x^\alpha, w^i, w_6\}$ (and $w_5=0$) we find, for massless
light quark such that $d=0$, 
\begin{equation} \label{action} 
  S_P = - \frac{T}{2} \int d \tau \int\limits_0^D d \sigma \left[
- \frac{\rho^2 + \sigma^2}{R^2} \dot x^\alpha \dot x_\alpha
 - \frac{R^2}{(\rho^2 + \sigma^2)}
\dot w^i\dot w_i + \frac{R^2}{(\rho^2 + \sigma^2)}
\right] .  
\end{equation}

Here we have used the background AdS metric in the conventions of
(\ref{AdSmetric})  with $\rho^2 = w^i w_i$, $i=1,2,3,4$.
Integrating (\ref{action}) over
$\sigma$ gives 
\begin{equation} 
  S_P = - \frac{T}{2} \int d \tau \left[-
f(\rho) \dot{x}^2 - g(\rho) \dot{w}_i^2 + g(\rho) \right] ,
\label{poly}
\end{equation} 
where 
\begin{equation} \label{adsfg} 
  f(\rho) = R^{-2} [\rho^2 D +\frac{1}{3} D^3]\,, \hspace{1cm} 
  g(\rho) =\frac{R^2}{\rho} \arctan (D/\rho) \, . 
\end{equation}

We wish to convert this classical dynamics to a second quantised
field that will play the role of the holographic field to the
heavy-light quark operators. We need the modified energy momentum
relation for the particle. That relation is provided by the
constraint equations. Firstly note that
\begin{equation} 
  \dot{X}^\mu X'_\mu = 0 
\end{equation}
is trivially satisfied because the string is not allowed to move in the $w_6$
direction. The remaining constraint, when integrated over $\sigma$, gives
\begin{equation} \label{Legendre} 
  \frac{1}{ f(\rho)} p_{x}^2 +\frac{1}{
g(\rho)} p_{w}^2 + T^2 g(\rho) =0 \, , 
\end{equation} 
where $p_x{}^\alpha \equiv \delta {\cal L} / \delta \dot x_\alpha$, 
$p_w{}^i \equiv  \delta {\cal L} / \delta \dot w_i$. 
Note that (\ref{Legendre}) is a simple modification of the usual
$E^2-p^2=m^2$ with the effective mass depending on the $\rho$ position
of the string. It is worth stressing the form of this
  energy-momentum relation at large radius $\rho$. Expanding
  (\ref{adsfg}), we see that in the UV limit, $f g$ becomes a constant
  while $f/g$ blows up as $\rho^4$. In this limit, (\ref{Legendre})
  becomes
\begin{equation} \label{eomrho}
  p_x^2 + \frac{\rho^4}{R^4} \,  p_w^2 + T^2 D^2 = 0 \,.
\end{equation}
$T^2 D^2$ is just the heavy quark
mass squared.  Note that for motion in the $x$ directions the string
behaves, as one might naively expect, as a massive string of mass
$m_H$. However for motion in the holographic $w$ directions the mass
of the string is $\rho$ dependent, and the string is effectively
massless at large $\rho$. The form of this asymptotic equation is 
easily understood in terms of the field theory dilations 
(see our discussion under (\ref{AdSmetric}) above)
-- the $w_i$ (as well as $\rho$) scale as  field theory energies,
such that the canonical momenta $p_w$ scale as a  length from the point of
view of the field theory. Thus the factor of $\rho^4$ in must be
present in (\ref{eomrho}) 
to match the dimension of $p_x^2$. (Note that $R$ does not
scale under four-dimensional conformal transformations.)

Making the usual quantum mechanical operator substitutions, we
arrive at the field equation
\begin{equation} 
  \left[\Box_x^2 +\frac{f(\rho)}{ g(\rho)} \nabla_w^2 - T^2  
f(\rho) g(\rho)  \right] \phi = 0 \, .\label{fieldequation}
\end{equation}
This is a modified Klein-Gordon equation. 
In the UV limit $\rho \rightarrow \infty$, we have
\begin{equation}  \label{fe2}
 \nabla_w^2 \, \phi = 0\, .
\end{equation}
(\ref{fe2}) is the four-dimensional Laplace equation and has solutions
of the form
\begin{equation} 
  \phi = m_{HL} + \frac{c_{HL}}{\rho^2}+\dots \, .
\end{equation}
This is the correct form to describe the source and \acro{VEV} of the
heavy-light operator $\bar \psi_{{H}}\psi_L$.

\begin{figure}
\begin{center}
\begin{psfrags}\input{hl-ads-converge-psfrag}\includegraphics[width=12.5cm,trim=40 0 0 0]{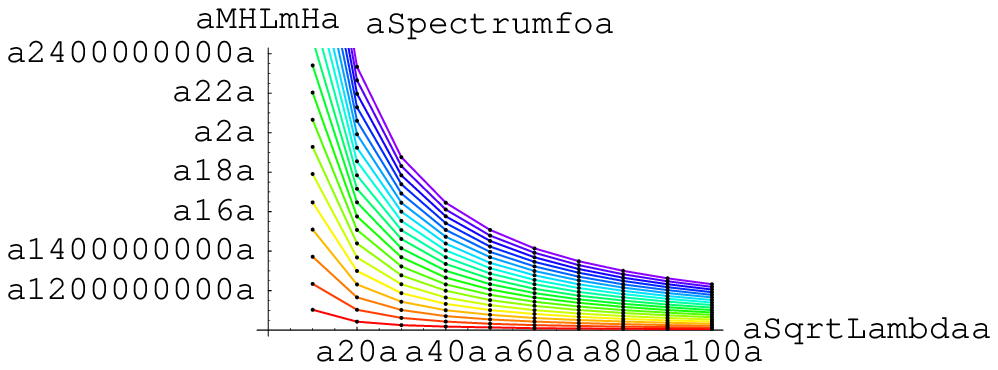}\end{psfrags}
\begin{psfrags}\input{hl-ads-linear-psfrag}\includegraphics[width=12.5cm]{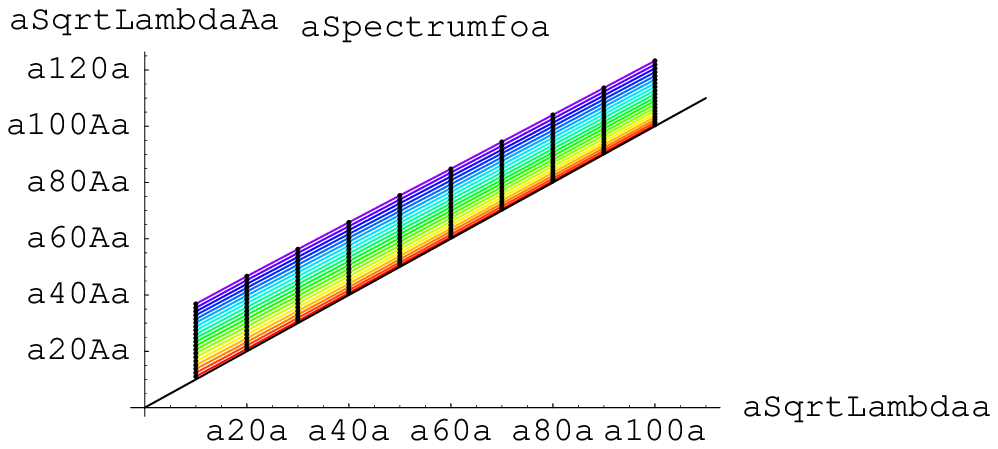}\end{psfrags}
\caption{The masses $M_{HL}$ of the meson and its excited states for
  the AdS background.
 The ratio $M_{HL}/m_H$, with $m_H$ the heavy quark mass 
(the light quark is taken
to be massless), is plotted against the square root of the 't~Hooft
coupling  $\lambda$.
 We observe that in the large $\lambda$ limit, $M_{HL}/m_H$
  behaves as $1+\text{const}/{\sqrt \lambda} +
  \mathcal{O}(\lambda^{-1})$. The black line in the second plot
  corresponds to $M_{HL}/m_H = 1$.\label{f:ads1}}
\end{center}
\end{figure}

\begin{figure}
\begin{center}
\begin{psfrags}\input{hl-ads-small-psfrag}\includegraphics[width=15cm]{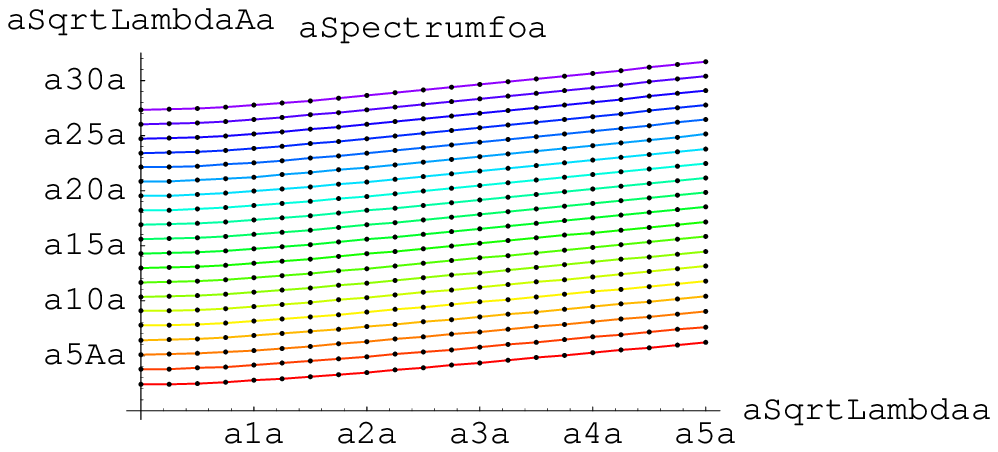}\end{psfrags}
\end{center}
\caption{A plot of $M_{HL}/m_H$ as function of the 't~Hooft coupling $\lambda$
for small 't~Hooft coupling (the light quark mass is zero) in AdS\@.
For small values of $\lambda$, one obtains $M_{HL}/m_H \sim
  \text{const.}/\sqrt{\lambda} + \mathcal{O}(\lambda)$. Note however
  that the supergravity background is unreliable in this
  limit.\label{f:ads2}}
\end{figure}

Since there is no heavy-light mass mixing term and no heavy-light
bilinear \acro{VEV} in \acro{QCD}, the correct background configuration has
$m_{HL},c_{HL}=0$ and we look at linearized  fluctuations of
the form
\begin{equation} \label{ansatz} \phi = h(\rho) e^{ik \cdot x},
  \hspace{1cm} M_{HL}{}^2= -k^2 \, ,
\end{equation} 
where the large $\rho$ behaviour of $h$ is $1/\rho^2$. We substitute
the ansatz (\ref{ansatz}) into (\ref{fieldequation}) and search
numerically for regular solutions. It is most convenient to do so in
rescaled coordinates $\varrho= \rho/D$, such that equation
\eqref{fieldequation} for our ansatz takes the form
\begin{align} 
  \Biggl\{\frac{\pi}{\lambda} \frac{ \varrho^3 + \frac{\varrho}{3} }{ \arctan \frac{1}{\varrho} } \nabla_\varrho^2 
    + \biggl[ \varrho + \frac{1}{3\varrho} \biggr] \arctan \frac{1}{\varrho}
    + \bar{M}_{HL}{}^2 \Biggr\} h (\varrho) = 0 \, , \label{adsnumerical}\qquad
   \bar{M}_{HL}{}^2&=\frac{M_{HL}{}^2}{m_H^2} \, .
\end{align}
By finding solutions for which $h$ is regular, this equation allows us
to calculate the ratio $M_{HL}/ m_H$ as a function of the 't Hooft
coupling $\lambda$. Note that from the standard AdS/CFT relation $R^4
= 4 \pi g_s N \alpha'{}^2$ we have $R^2T = \sqrt{\lambda/\pi}$. 
The results for the masses of the meson and its excited states
are shown in figures \ref{f:ads1} and \ref{f:ads2}.
We read off that $M_{HL}/m_H = 1 + \text{const}/{\sqrt \lambda} +
\mathcal{O}(\lambda^{-1})$. We see that in the large $\lambda$ limit
we have $M_{HL}=m_H$ (this follows since in this limit the $\varrho$ 
derivative in (\ref{adsnumerical}) is suppressed).
This is in agreement with the naive expectation that the meson mass should be 
equal to string length times its tension.

For comparison we also plot the meson mass dependence on the 't Hooft
coupling for small values of $\lambda$ in figure \ref{f:ads2}. Note
however that the supergravity approximation is unreliable in this
limit. 

%




\newpage

\section{Heavy-Light Mesons in Dilaton Flow Geometry}

\subsection{Dilaton Flow Geometry}

The AdS geometry studied so far does not contain the physics of most
interest in the heavy-light sector of \acro{QCD}. The main interest
lies in the case where the light quark is involved in chiral symmetry
breaking and has a dynamically generated mass whilst the heavy quark
does not. Chiral symmetry breaking is forbidden in a supersymmetric
theory so we will turn instead to a non-supersymmetric dilaton flow
geometry that provides a gravity dual of \acro{QCD} \cite{BEEGK}.

That model consists of a  deformed AdS geometry
\cite{CM} -- \cite{Gubser2}
\begin{equation} 
  ds^2 = H^{-1/2} \left( \frac{ w^4 + b^4}{w^4-b^4}
\right)^{\delta/4} dx_{4}^2 + H^{1/2} \left( \frac{w^4 + b^4}{ w^4-
b^4}\right)^{(2-\delta)/4} \frac{w^4 - b^4}{w^4} \sum_{i=1}^6
dw_i^2, 
\end{equation} 
where
\begin{equation} 
  H =  \left(  \frac{ w^4 + b^4}{w^4 - b^4}\right)^{\delta} - 1
\end{equation} 
and the dilaton and four-form are given by
\begin{equation} 
  e^{2 \phi} = e^{2 \phi_0} \left( \frac{ w^4 + b^4}{w^4 - b^4} \right)^{\Delta}, \hspace{1cm} 
  C_{(4)} = - \frac{1}{4} H^{-1} dt \wedge dx \wedge dy \wedge dz \, . 
\end{equation} 
There are formally two free parameters, $R$ and $b$, since
\begin{equation} 
  \delta = \frac{R^{4}}{2 b^4}\,,\hspace{1cm} 
  \Delta^2 = 10 - \delta^2 \, .
\end{equation}

We see that $b$ has length dimension one and enters to the fourth
power. The SO(6) symmetry of the geometry is retained at all
values of the radius, $w$. We conclude that in the field theory a
dimension four operator with no SO(6) charge has been switched on.
$b^4/ (2\pi \alpha')^4$ therefore corresponds to a \acro{VEV} for the
operator $\Tr F^2$. Since $b$ is the only conformal symmetry
breaking parameter in the theory it sets the intrinsic scale of
strong coupling in terms of $\Lambda_b = b /(2\pi \alpha')$. In
order to present numerical results below, we will set $b=R$ which
corresponds to $\delta=1/2$ -- this is a representative point in the
parameter space that displays chiral symmetry breaking.


Quarks are again introduced by including probe D7 branes into the
geometry. Substituting into the DBI action (\ref{DBI}) for this
geometry we find \cite{BEEGK} the equation of motion for the radial
separation, $v$, of the two branes in the $8,9$ directions as a
function of the radial coordinate $\rho^2= \sum_{i=1}^4 w_i^2$ in the
$4-7$ directions,
\begin{equation} \label{eommc}
  \frac{ d}{d \rho} \left[ \frac{e^{\phi}  { \cal
  G}(\rho,v)}{\sqrt{ 1 + (\partial_\rho v)^2}  }
  (\partial_\rho v)\right] - \sqrt{ 1 + \partial_\rho v^2}
  \frac{d}{d {v}} \left[ e^{\phi} { \cal G}(\rho,v)
  \right] = 0 \, ,
\end{equation} 
where 
\begin{equation}
  {\cal G}(\rho,v) =  \rho^3 \frac{( (\rho^2 + v^{2})^2 +
  b^4) ( (\rho^2 + v^{2})^2 - b^4)}{(\rho^2 + v^{2})^4}   \, .
\end{equation} 
At large $\rho$ the solutions take the form \begin{equation}
  v = m + \frac{c}{\rho^2} + \dots \, . \label{asymp}
\end{equation}

\begin{figure}[t]
\begin{center}
  \includegraphics[height=6cm]{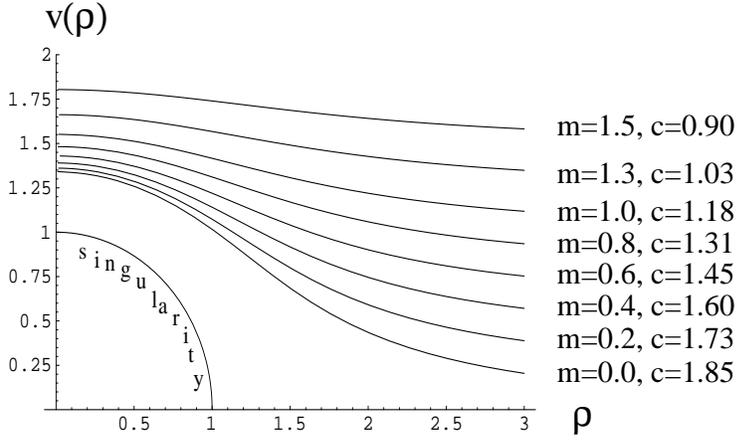}
  \caption{A plot of the embedding of the D7 brane as a function of the
  radial coordinate $\rho$.}\label{f:embed}
\end{center}
\end{figure}

There are two free parameters in the solution. The first, $m$,
describes the asymptotic separation of the D3 and D7 branes and has
conformal dimension one -- it is the bare quark mass. The second
parameter, $c$, has dimension three and corresponds to the $\bar{\psi}
\psi$ quark bilinear. To obtain solutions which are regular in the \acro{IR},
we impose the boundary condition $\dot{v}(0) = 0$, where the dot
indicates a $\rho$ derivative, as well as fixing $m$ in the \acro{UV}.
Regular solutions are displayed in figure \ref{f:embed}. The
regularity condition fixes the condensate $c$ as a function of the
quark mass $m$.

The solutions show that a dynamical mass is formed for the quarks.
A massless quark would correspond to a D7 brane that intersects
the D3 brane, such that there was a zero length string between them. We
see that the D3's repel the D7 and for all configurations there is
a non-zero minimum length string. The solution which
asymptotically has $m=0$ also explicitly breaks the $\mathrm{U}(1)$ symmetry
in the $8,9$ plane by bending off the axis. This is the geometric
representation of the spontaneous breaking of the $\mathrm{U}(1)$ axial symmetry of the
quarks.

Fluctuations of the brane about the solution found above in the $8,9$
directions correspond to excitations of the operator $\bar{\psi} \psi$
and contain information about the pion and sigma field of the model.
With $w_6+iw_5 = v(\rho)U(\rho,x)$ and expanding to second order in
$U(\rho,x)$, the DBI action (\ref{DBI}) becomes
\begin{equation}
  S=-T_7 \int d^8\xi~e^\phi\sqrt{-g}(1+\dot v^2)^{\frac{1}{2}}
  \left[1+\frac{1}{2} g_{\rho\rho} v^2(1+\dot v^2)^{-1}
  \partial^aU\partial_aU^\dag\right]\, .
\end{equation} 
With $U(\rho,x)=\exp(i\pi(\rho,x))$, this gives an action for the pion
field $\pi$ and for the sigma field (here denoted by $v$).

There is also a superpartner U(1) gauge field in the action which
describes the operator $\bar{\psi} \gamma^\mu \psi$ and hence vector
mesons. This is introduced as a gauge field $F_{ab}$ living on the D7,
such that the DBI action now reads
\begin{equation}
  S_F=-T_7 \int d^8\xi~e^\phi\left[-\det({\bf P}[g_{ab}]+
  2\pi\alpha'e^{-\phi/2}F_{ab})\right]^{\frac{1}{2}},
\end{equation} 
which, expanded to second order, gives 
\begin{equation}
  S_F=-T_7  \int d^8\xi~e^\phi\sqrt{-g}(1+\dot v^2)^{\frac{1}{2}}
  \left[1+\frac{1}{2} g_{\rho\rho} v^2(1+\dot v^2)^{-1}
  \partial^aU\partial_aU^\dag - \frac{1}{4}(2\pi\alpha')^2 e^{-\phi}F^2\right].
\end{equation}

The pseudo-scalar mesons, i.e.~the pions, correspond to fluctuations
in the position of the brane in the angular direction in the $w_5-w_6$
plane (we have suppressed fluctuations of the radial field $v$ for
simplicity). In this paper we use the predictions for the vector meson
masses. They correspond to regular solutions of the equation of motion
for the gauge potential $A_\mu$. The equation of motion for a solution
of the form $A_\mu = V_\mu(\varrho) e^{iq\cdot x}, M^2=-q^2$ is (we use the scaled 
coordinate $\hat{\rho} = \rho/b,~ \hat{v}=v/b$)
\begin{equation}\label{eq:vec-eom-2}
   \partial_{\hat{\rho}}(K_1(\hat{\rho})\partial_{\hat{\rho}} V_\mu(\hat{\rho}))
-b^2 q^2K_2(\hat{\rho})V_\mu(\hat{\rho})=0 \, ,
\end{equation} with
\begin{equation*}
K_1=X^{1/2}Y \hat{\rho}^3(1+\dot{\hat{v}}^2)^{-1/2},~~ K_2= H
X^{1-\delta/2}Y^2\hat{\rho}^3(1+\dot{\hat{v}}^2)^{-1/2}  \, ,
\end{equation*}
and
\begin{equation}\label{eq:XYdef}
  X =
  \frac{(\hat{v}^2+\hat{\rho}^2)^2+1}{(\hat{v}^2+\hat{\rho}^2)^2-1}
 \,  , ~~~~~ Y = \frac{(\hat{v}^2+\hat{\rho}^2)^2-1}
{(\hat{v}^2+\hat{\rho}^2)^2} \, .
\end{equation}

The boundary conditions imply $V^\mu \sim 1/\hat{\rho}^2$ at large
$\hat{\rho}$.  
Note that in the coordinates we are using, (\ref{eq:vec-eom-2}) 
provides us with the quantity $b M_{vec}=RM_{vec}/(2 \delta)^{1/4}$, with $M_{vec}$ the meson mass,
as a function of the field theory quantities
$m_{\text{quark}}/\Lambda_b$ with $\Lambda_b=bT$ being the effective \acro{QCD} scale.

\subsection{Heavy-Light Mesons}

The formalism for studying heavy-light mesons as presented in
section \ref{hlm}  moves across almost directly from the AdS
geometry to the dilaton flow case. Since the five-sphere is
undeformed, the action for a string tied between two D7 branes
again takes the form 
\begin{equation} S_P = - \frac{T }{2} \int d \tau
\left[- f(\rho) \dot{x}^2 - g(\rho) \dot{w}_i^2 + g(\rho) \right]
\,
\end{equation} 
as in (\ref{poly}),   where now
\begin{align}\label{cmfg} 
\begin{aligned}
  f(\rho)
      &= \int\limits_{v(m_L)}^{v(m_H)} dv\,e^{\phi/2} g_{xx}
       = \int\limits_{v(m_L)}^{v(m_H)} dv\,
         (X^{1/2}-1)^{-1/2} X^{(\Delta+\delta)/4} \, , \\
  g(\rho)
      &=  \int\limits_{v(m_L)}^{v(m_H)} dv\, e^{\phi/2} g_{ww}
       =  \int\limits_{v(m_L)}^{v(m_H)} dv\,
          (1-b^4(v^2+\rho^2)^{-2}) (X^\delta-1)^{1/2}
          X^{(2+\Delta-\delta)/4} \, ,
\end{aligned}
\end{align} 
with $X$ defined in \eqref{eq:XYdef}.  The integration limits for the
$v$ integrals, $v(m_h)$ and $v(m_l)$, are given by the D7 probe
embeddings found above, as shown in \ref{f:embed}. Since we only know
those embeddings numerically, $f$ and $g$ are also only known
numerically.  Nevertheless we find the regular solutions of the wave
equation
\begin{equation} \label{hlcm}
\left[\Box_x^2 +\frac{f(\rho)}{ g(\rho)} \nabla_w^2 - T^2 f(\rho) g(\rho)
 \right] \phi = 0 \, , 
\end{equation} 
again using numerics. 
The \acro{UV} limit coincides with the AdS model with solutions $ \phi =
m_{HL} + \frac{c_{HL}}{\rho^2}+\cdots$. We take $m_{HL}$,
$c_{HL}=0$ since such mass mixing and condensate terms are absent
in \acro{QCD}. Then we again seek linearized fluctuations of the form
\begin{equation} 
  \phi = h(\rho) e^{ik\cdot x}, \hspace{1cm} M^2=-k^2 \, ,
\end{equation}
where the large $\rho$ behaviour of $h$ is $1/\rho^2$. To do so it
is again convenient to work
in rescaled coordinates, $\hat\rho=\rho/b$ and $\hat{v}=v/b$, such that
equation \eqref{hlcm} reads
\begin{align} \label{hlcm2}
  [ \frac{2\pi\delta}{\lambda} \hat{f}/\hat{g} \nabla_{\hat{\rho}}^2  
    + \frac{M^2}{\Lambda_b^2} - \hat{f} \hat{g} ] \phi(\hat{\rho}) = 0
    \, ,
\end{align}
where $\Lambda_b=bT$ is the \acro{QCD} scale; while $\hat{f}$ and $\hat{g}$ are dimensionless quantities that can be
effectively obtained from \eqref{cmfg} by setting $b=1$. In these coordinates
the heavy quark mass enters through the integrations limits in \eqref{cmfg} in the combination $m_H / \Lambda_b$.
\begin{figure}
\begin{center}
\begin{psfrags}\input{hl-cm-xmas-compare-lambda50-all-psfrag}\includegraphics[width=14cm]{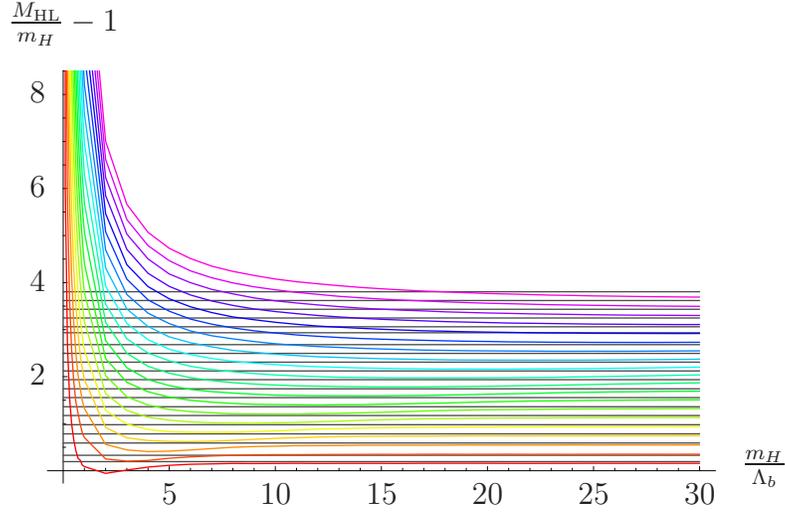}\end{psfrags}
\caption{Binding energy as a function of the ratio heavy quark mass
  over QCD scale for the Constable-Myers background, for 't~Hooft
  coupling $\lambda =10^4$.  For large ratio,
  the AdS behaviour (shown in black) is approached, whereas for small
  ratio effects of the chiral symmetry breaking are seen.   
}\label{cmmesons}
\end{center}
\end{figure}
\begin{figure}
\begin{center}
\begin{psfrags}\input{hl-comparison-psfrag}\includegraphics[width=14cm]{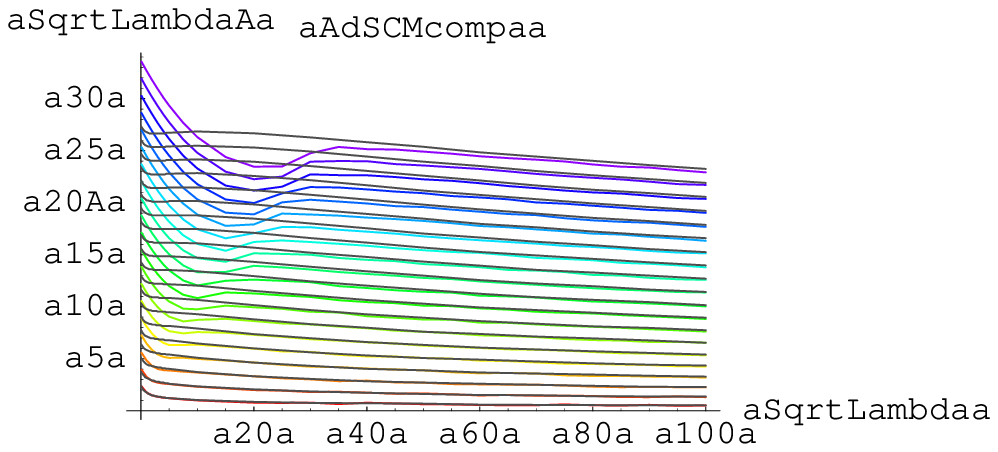}\end{psfrags}
\caption{Binding energy for fixed
$m_H/\Lambda_b = 12.63$ as a function of $\sqrt{\lambda}$ in the
Constable-Myers background. The AdS
behaviour is again shown in black for comparison.}

\label{cmmesons2}
\end{center}
\end{figure}

The masses of the heavy-light mesons are very similar in
this model to those in AdS. To make the deviations clear we
plot the binding energy $(M_{HL}-m_H)/m_H$. This quantity is shown in
figure 6
as a function of the heavy quark mass (with massless light
quark) and at fixed 't~Hooft coupling ($\sqrt{\lambda} =
100$). We find the same  basic
behaviour for all $\lambda$. When $m_H/\Lambda_b$ is large,
the model returns to AdS-like results. As the heavy
quark mass comes down towards the scale of the chiral
symmetry breaking $\Lambda_b$, there are larger deviations
as one would expect. Note that at large $m_H$ the binding
energy asymptotes to a constant -- this means that since all
the excited states have a mass which is to first order just
$m_H$, the percentage difference in the masses of these
states tends to zero in this limit. 

In figure 7 the binding energy is shown for a fixed
value of $m_H/\Lambda_b= 12.63$ as a function of $\lambda$. Again
the precise value of $m_H$ is not important here (though it will be
below). At large
$\lambda$ we again see convergence to the AdS results -- in
this limit the UV of the theory is very strongly coupled
and the growth of the coupling near $r=b$ (where it
diverges) is presumably less important than at small
$\lambda$, where the strongly coupled IR behaviour takes
more prominence. The higher excited states feel the effects
of the IR dynamics more strongly possibly because these
states are larger.

\section{Bottom Phenomenology}

Recently there have been a number of attempts to convert these stringy
holographic descriptions of chiral symmetry breaking into
phenomenological models of \acro{QCD}. In particular, models involving
a simple AdS slice \cite{Erlich:2005qh, DaRold:2005zs}, or the dilaton
flow geometry described above \cite{Evans:2006dj}, have proven to be
successful, giving agreement to light quark meson data at better than
the 20\% level. In many ways this good agreement with experiment is
surprising since these models become strongly coupled conformal
theories in the ultra-violet rather than asymptotically free.
Encouraged by the success of those models though let us consider
comparing our dilaton flow model, that incorporates chiral symmetry
breaking, to the bottom quark sector of the \acro{QCD} spectrum. The
D7 brane generating the bottom quark will lie in the AdS-like regime
of our model. In \acro{QCD} the dynamics should be perturbative at
this energy scale but this clearly cannot be achieved in a gravity
dual. We hope that AdS is the next best scenario since the quarks lie
in the conformal gauge background of the supersymmetric theory where
strong non-renormalization theorems apply, in some way mimicking a
perturbative regime.

In fact the models discussed in this paper appear somewhat different from
  \acro{QCD}.  The masses of the heavy-heavy and light-light meson
  states are suppressed relative to the quark mass by a factor of the
  't~Hooft coupling (see (\ref{llmass})) which according to the
  AdS/CFT Maldacena limit is big.  
On the other hand, the heavy-light meson 
masses are not suppressed by $\lambda$ (see fig.~3). 
A large parametric suppression of this form
is not apparent in the QCD data for the bottom quark although the $\Upsilon$ mass is less than 
twice the B meson mass ($m_\Upsilon = 1.8 M_B$).

Nevertheless we can attempt phenomenology. We use the equations for the vector meson
masses, $M_{vec}$, in the Constable-Myers geometry (\ref{eq:vec-eom-2})
to determine the value of the heavy (bottom) quark mass. In fact that equation gives 
$b M_{vec}$ as a function of  $m_H/\Lambda_b$,
so by engineering the correct ratio of the
$\Upsilon$ to $\rho$ meson masses (we assume the light quarks are massless) we find
$m_H / \Lambda_b = 12.63$. We then substitute this value of $m_H$ into the equations
for the heavy-light meson mass (\ref{hlcm2}). We still have the value
of the 
't~Hooft
parameter to fix which we can determine from requiring that we obtain the correct B-meson mass.

We can determine $\lambda$ from the physical value of the $B$ and
$\rho$ quark ratio using
\begin{equation}  \label{mb} 
\left(\frac{M_B}{M_\rho}\right)^{phys} = 
\frac{M_{HL}}{\Lambda_b}~\frac{\Lambda_b}{M_{{\text{vec}}, \, LL}} =
\frac{M_{HL}}{\Lambda_b} ~ \frac{1}{b M_{vec, \, LL}}~ \sqrt{\frac{2 \pi
    \delta}{\lambda}} \, .
\end{equation}
Calculating $M_{HL}/\Lambda_b$ as explained in section 3.2
for $m_H / \Lambda_b = 12.63$, the
relation (\ref{mb}) allows us to determine $\lambda$. We find
 $\lambda=5.22$. 
Whilst this value is not very large it  hopefully is sufficiently big
to make use of our large $\lambda$ approximation. 
Moreover, evaluating $M_{HL}/\Lambda_b$ 
at $\lambda = 5.22$, and identifying $M_{HL}$ with the physical $B$
quark mass $M_B = 5279$ MeV, we find for the QCD scale $\Lambda_b =
340$ MeV, which is a little too high compared with QCD expectations,
though  of the right order of magnitude. We also find $m_H = 12.63
\Lambda_b = 4294$ MeV for the physical $b$ quark mass.

\begin{figure}
\begin{psfrags}\input{hl-b-bstar-ratio-psfrag}\includegraphics[width=15cm]{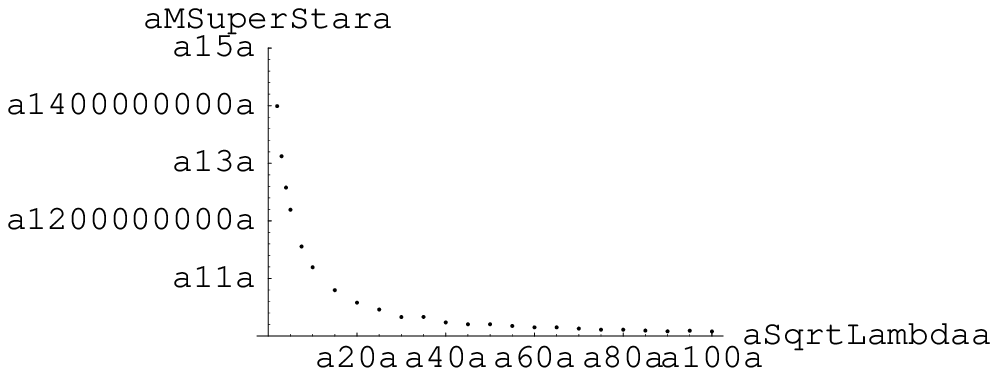}\end{psfrags}
\caption{Ratio of the masses of the first excited and lowest
  heavy-light states as a function of the 't~Hooft coupling.
From experimental data we expect $M_{B^*}/M_B \approx 1.01$.
In our model, this ratio converges to one for infinite $\lambda$.}

\label{f:ratio}
\end{figure}

We can now predict the excited B-meson masses. 
We plot $M_{B^*}/M_B$ against the 't~Hooft coupling for  $m_H/\Lambda_b
= 12.63$ in figure \ref{f:ratio}.
Formally we have in the spirit of (\ref{mb})
\begin{equation} M_{B^*}^{phys} = M_\rho^{phys}~
  \frac{M_B}{\Lambda_b}~ 
 \frac{1}{b M_\rho} ~
\sqrt{\frac{2 \pi \delta}{\lambda}} \, , \end{equation}
with all quantities on the right again computable at a given $\lambda$.
At $\lambda=5.2$ we predict $M_{B^*}=6403$ MeV which is 20\% larger 
than the measured value of $5325$ MeV. To approach the physical 
$M_{B^*}/M_B$ ratio (which is 1.01) would require a much
larger value of $\lambda$,  as can be seen in figure \ref{f:ratio}
 -- but in this case, the light-light and heavy-heavy vector meson
 masses would then 
be overly suppressed.\footnote{For comparison with \acro{QCD} phenomenology, we note that 
there are flux tube and \acro{QCD} string models which allow the
calculation of the B mass from the $\rho$ and $\Upsilon$ masses, for
instance \cite{Kaidalov,Simonov}.} 

In conclusion, although the model is not a perfect description of QCD, 
it does display the
approximate pattern of the QCD b-quark mesons if we take a moderate value
of the 't~Hooft coupling.

\bigskip \bigskip
\bigskip \bigskip

\noindent {\bf Acknowledgements:} The authors are grateful to
R.~Apreda, A.~Hoang, I.~Kirsch, C.~Sieg and to T.~Waterson for
discussions, as well as to M.~Shifman for pointing out references
\cite{Kaidalov,Simonov}.  Moreover they are grateful to R.~Rashkov and
to J.~Shock for comments on the first version of the paper.

\bigskip \bigskip

\bigskip \bigskip

\bigskip \bigskip

\bigskip \bigskip

\bigskip \bigskip



\begin{thebibliography}{ll}





\bibitem{Mal}
J.~M.~Maldacena,  Adv.\ Theor.\ Math.\ Phys.\  {\bf 2}, 231 (1998)
Int.\ J.\ Theor.\ Phys.\  {\bf 38}, 1113 (1999)
[arXiv:hep-th/9711200].

\bibitem{Gub}
S.~S.~Gubser, I.~R.~Klebanov and A.~M.~Polyakov,  Phys.\ Lett.\ B
{\bf 428}, 105 (1998) [arXiv:hep-th/9802109].

\bibitem{Wit}
E.~Witten,  Adv.\ Theor.\ Math.\ Phys.\  {\bf 2}, 253 (1998)
[arXiv:hep-th/9802150].



\bibitem{BEEGK}
  J.~Babington, J.~Erdmenger, N.~J.~Evans, Z.~Guralnik and I.~Kirsch,
  Phys.\ Rev.\ D {\bf 69} (2004) 066007
  [arXiv:hep-th/0306018].



\bibitem{Evans:2004ia}
  N.~J.~Evans and J.~P.~Shock,
  Phys.\ Rev.\ D {\bf 70} (2004) 046002
  [arXiv:hep-th/0403279].




\bibitem{Evans:2005ti}
  N.~Evans, J.~Shock and T.~Waterson,
  JHEP {\bf 0503} (2005) 005
  [arXiv:hep-th/0502091].




\bibitem{Apreda:2005hj}
  R.~Apreda, J.~Erdmenger and N.~Evans,
JHEP {\bf 0605} (2006) 011 [arXiv:hep-th/0509219].

\bibitem{Apreda:2005yz}
  R.~Apreda, J.~Erdmenger, N.~Evans and Z.~Guralnik,
  Phys.\ Rev.\ D {\bf 71} (2005) 126002
  [arXiv:hep-th/0504151].


\bibitem{Mateos2}
M.~Kruczenski, D.~Mateos, R.~C.~Myers and D.~J.~Winters,
 JHEP {\bf 0405} (2004) 041
[arXiv:hep-th/0311270].

\bibitem{Barbon:2004dq}
  J.~L.~F.~Barbon, C.~Hoyos, D.~Mateos and R.~C.~Myers,
  JHEP {\bf 0410} (2004) 029
  [arXiv:hep-th/0404260].



\bibitem{Ghoroku:2004sp}
  K.~Ghoroku and M.~Yahiro,
  Phys.\ Lett.\ B {\bf 604} (2004) 235
  [arXiv:hep-th/0408040].



\bibitem{Brevik:2005fs}
  I.~Brevik, K.~Ghoroku and A.~Nakamura,
  Int.\ J.\ Mod.\ Phys.\ D {\bf 15} (2006) 57
  [arXiv:hep-th/0505057].


\bibitem{Sakai:2004cn}
  T.~Sakai and S.~Sugimoto,
  Prog.\ Theor.\ Phys.\  {\bf 113} (2005) 843
  [arXiv:hep-th/0412141].




\bibitem{Sakai:2005yt}
  T.~Sakai and S.~Sugimoto,
  Prog.\ Theor.\ Phys.\  {\bf 114} (2006) 1083
  [arXiv:hep-th/0507073].




\bibitem{Antonyan:2006vw}
  E.~Antonyan, J.~A.~Harvey, S.~Jensen and D.~Kutasov,
  arXiv:hep-th/0604017.

\bibitem{Bak:2004nt}
  D.~Bak and H.~U.~Yee,
  Phys.\ Rev.\ D {\bf 71} (2005) 046003
  [arXiv:hep-th/0412170].




\bibitem{Ghoroku:2005tf}
  K.~Ghoroku, T.~Sakaguchi, N.~Uekusa and M.~Yahiro,
  Phys.\ Rev.\ D {\bf 71} (2005) 106002
  [arXiv:hep-th/0502088].



\bibitem{Mateos:2006nu}
  D.~Mateos, R.~C.~Myers and R.~M.~Thomson,
   Phys.\ Rev.\ Lett.\  {\bf 97} (2006) 091601 
[arXiv:hep-th/0605046].



\bibitem{Albash:2006ew}
  T.~Albash, V.~Filev, C.~V.~Johnson and A.~Kundu,
  arXiv:hep-th/0605088.

\bibitem{Albash:2006bs}
  T.~Albash, V.~Filev, C.~V.~Johnson and A.~Kundu,
  arXiv:hep-th/0605175.




\bibitem{Parnachev:2006dn}
  A.~Parnachev and D.~A.~Sahakyan,
Phys.\ Rev.\ Lett.\  {\bf 97} (2006) 111601
  [arXiv:hep-th/0604173].

\bibitem{Aharony:2006da}
  O.~Aharony, J.~Sonnenschein and S.~Yankielowicz,
  arXiv:hep-th/0604161.




\bibitem{Karch:2006bv}
  A.~Karch and A.~O'Bannon,
 Phys.\ Rev.\ D {\bf 74} (2006) 085033  [arXiv:hep-th/0605120].



\bibitem{Erlich:2005qh}
  J.~Erlich, E.~Katz, D.~T.~Son and M.~A.~Stephanov,
  Phys.\ Rev.\ Lett.\  {\bf 95} (2005) 261602
  [arXiv:hep-ph/0501128].

\bibitem{DaRold:2005zs}
  L.~Da Rold and A.~Pomarol,
  Nucl.\ Phys.\ B {\bf 721} (2005) 79
  [arXiv:hep-ph/0501218].

\bibitem{deTeramond:2005su}
  G.~F.~de Teramond and S.~J.~Brodsky,
  Phys.\ Rev.\ Lett.\  {\bf 94} (2005) 201601
  [arXiv:hep-th/0501022].


\bibitem{Brodsky:2006uq}
  S.~J.~Brodsky and G.~F.~de Teramond,
  Phys.\ Rev.\ Lett.\  {\bf 96} (2006) 201601  [arXiv:hep-ph/0602252].

\bibitem{Hirn:2005nr}
  J.~Hirn and V.~Sanz,
  JHEP {\bf 0512} (2005) 030
  [arXiv:hep-ph/0507049].

\bibitem{Hirn:2005vk}
  J.~Hirn, N.~Rius and V.~Sanz,
  Phys.\ Rev.\ D {\bf 73} (2006) 085005
  [arXiv:hep-ph/0512240].



\bibitem{Boschi-Filho:2002ta}
  H.~Boschi-Filho and N.~R.~F.~Braga,
  Eur.\ Phys.\ J.\ C {\bf 32} (2004) 529
  [arXiv:hep-th/0209080].

\bibitem{Boschi-Filho:2002vd}
  H.~Boschi-Filho and N.~R.~F.~Braga,
  JHEP {\bf 0305} (2003) 009
  [arXiv:hep-th/0212207].


\bibitem{Hong:2003jm}
  S.~Hong, S.~Yoon and M.~J.~Strassler,
  JHEP {\bf 0404} (2004) 046
  [arXiv:hep-th/0312071].




\bibitem{Evans:2006dj}
  N.~Evans and T.~Waterson,
  JHEP {\bf 0701} (2007) 058 [arXiv:hep-ph/0603249].

\bibitem{DaRold:2005vr}
  L.~Da Rold and A.~Pomarol,
  JHEP {\bf 0601} (2006) 157
  [arXiv:hep-ph/0510268].


  \bibitem{Shock:2006fc}
 J.~P.~Shock,
JHEP {\bf 0610} (2006) 043  [arXiv:hep-th/0601025].



\bibitem{Shock:2006qy}
  J.~P.~Shock and F.~Wu,
 JHEP {\bf 0608} (2006) 023  [arXiv:hep-ph/0603142].





\bibitem{Ghoroku:2005kg}
  K.~Ghoroku and M.~Yahiro,
 Phys.\ Rev.\ D {\bf 73} (2006) 125010  [arXiv:hep-ph/0512289].


\bibitem{Ghoroku:2005vt}
  K.~Ghoroku, N.~Maru, M.~Tachibana and M.~Yahiro,
  Phys.\ Lett.\ B {\bf 633} (2006) 602
  [arXiv:hep-ph/0510334].

\bibitem{Ghoroku:2006cc}
  K.~Ghoroku, A.~Nakamura and M.~Yahiro,
 Phys.\ Lett.\ B {\bf 638} (2006) 382 [arXiv:hep-ph/0605026].





\bibitem{Karch:2006pv}
  A.~Karch, E.~Katz, D.~T.~Son and M.~A.~Stephanov,
Phys.\ Rev.\ D {\bf 74} (2006) 015005  [arXiv:hep-ph/0602229].


  \bibitem{Andreev:2006vy}
  O.~Andreev,
 Phys.\ Rev.\ D {\bf 73} (2006) 107901  [arXiv:hep-th/0603170].

\bibitem{Andreev:2006ct}
  O.~Andreev and V.~I.~Zakharov,
Phys.\ Rev.\ D {\bf 74}, 025023 (2006)  [arXiv:hep-ph/0604204].




\bibitem{Hambye:2005up}
  T.~Hambye, B.~Hassanain, J.~March-Russell and M.~Schvellinger,
 Phys.\ Rev.\ D {\bf 74} (2006) 026003  [arXiv:hep-ph/0512089].






\bibitem{KarchKatz}
A.~Karch and E.~Katz,  JHEP {\bf 0206} (2002) 043
[arXiv:hep-th/0205236];

\bibitem{add1}  M. Bertolini, P. Di Vecchia, M. Frau, A. Lerda and R.
Marotta,  Nucl. Phys. {\bf B621}, 157, [arXiv:hep-th/0107057].

\bibitem{Polchinski}  M.
Gra\~{n}a and J. Polchinski, Phys. Rev. {\bf D65}:126005 (2002),
[arXiv:hep-th/0106014].


\bibitem{Mateos}
M.~Kruczenski, D.~Mateos, R.~C.~Myers and D.~J.~Winters, 
JHEP {\bf 0307} (2003) 049
[arXiv:hep-th/0304032].



\bibitem{Myers:2006qr}
  R.~C.~Myers and R.~M.~Thomson,
 JHEP {\bf 0609} (2006) 066
  [arXiv:hep-th/0605017]. 



\bibitem{Sonnenschein}
T.~Sakai and  J.~Sonnenschein, JHEP {\bf 0309} (2003) 047,
[arXiv:hep-th/0305049].

\bibitem{Erdmenger:2005bj}
  J.~Erdmenger, J.~Gro\ss e and Z.~Guralnik,
  JHEP {\bf 0506} (2005) 052
  [arXiv:hep-th/0502224].



\bibitem{Kirsch:2005uy}
  I.~Kirsch and D.~Vaman,
  Phys.\ Rev.\ D {\bf 72} (2005) 026007
  [arXiv:hep-th/0505164].



\bibitem{Arean:2006pk}
  D.~Arean and A.~V.~Ramallo,
   JHEP {\bf 0604} (2006) 037
  [arXiv:hep-th/0602174]. 






\bibitem{Nunez} C. Nunez, A. Paredes and A.V. Ramallo,
 JHEP
{\bf 0312} (2003) 024, [arXiv: hep-th/0311201].


\bibitem{Wang} X. Wang, S. Hu,
JHEP {\bf 0309} (2003) 017, [arXive: hep-th/0307218].

\bibitem{Erdmenger:2004dk}
  J.~Erdmenger and I.~Kirsch,
  JHEP {\bf 0412} (2004) 025
  [arXiv:hep-th/0408113].



\bibitem{Casero:2006pt}
  R.~Casero, C.~Nunez and A.~Paredes,
  Phys.\ Rev.\  D {\bf 73} (2006) 086005
  [arXiv:hep-th/0602027].

\bibitem{Edelstein:2006kw}
  J.~D.~Edelstein and R.~Portugues,
  Fortsch.\ Phys.\  {\bf 54}, 525 (2006)
  [arXiv:hep-th/0602021]. 

\bibitem{Casero:2005se}
  R.~Casero, A.~Paredes and J.~Sonnenschein,
  JHEP {\bf 0601} (2006) 127
  [arXiv:hep-th/0510110].


\bibitem{Hirayama:2006jn}
  T.~Hirayama,
 JHEP {\bf 0606}, 013 (2006)
  [arXiv:hep-th/0602258]. 

  \bibitem{Janik:2005zt}
  R.~A.~Janik and R.~Peschanski,
  Phys.\ Rev.\ D {\bf 73} (2006) 045013
  [arXiv:hep-th/0512162].

\bibitem{Canoura:2005uz}
  F.~Canoura, J.~D.~Edelstein, L.~A.~P.~Zayas, A.~V.~Ramallo and D.~Vaman,
  JHEP {\bf 0603} (2006) 101
  [arXiv:hep-th/0512087].

\bibitem{Benvenuti:2005qb}
  S.~Benvenuti, M.~Mahato, L.~A.~Pando Zayas and Y.~Tachikawa,
  arXiv:hep-th/0512061.


\bibitem{Paredes:2004is}
  A.~Paredes and P.~Talavera,
  Nucl.\ Phys.\ B {\bf 713} (2005) 438
  [arXiv:hep-th/0412260].


\bibitem{Pons:2004dk}
  J.~M.~Pons, J.~G.~Russo and P.~Talavera,
  Nucl.\ Phys.\ B {\bf 700} (2004) 71
  [arXiv:hep-th/0406266].



\bibitem{Peeters}
  K.~Peeters, J.~Sonnenschein and M.~Zamaklar,
  JHEP {\bf 0602} (2006) 009
  [arXiv:hep-th/0511044].



\bibitem{Troost}
  A.~L.~Cotrone, L.~Martucci and W.~Troost,
   Phys. \ Rev. \ Lett. \ {\bf 96} (2006) 141601 [arXiv:hep-th/0511045].



\bibitem{Cotrone:2006re}
  A.~L.~Cotrone, L.~Martucci, J.~M.~Pons and P.~Talavera,
  arXiv:hep-th/0604051.



\bibitem{Bando:2006sp}
  M.~Bando, A.~Sugamoto and S.~Terunuma,
Prog. \ Theor. \ Phys. \ {\bf 115} (2006) 1111  [arXiv:hep-ph/0602203].



\bibitem{Herzog:2006gh}
  C.~P.~Herzog, A.~Karch, P.~Kovtun, C.~Kozcaz and L.~G.~Yaffe,
  JHEP {\bf 0607} (2006) 013 [arXiv:hep-th/0605158].




\bibitem{Manohar}
  A.~V.~Manohar and M.~B.~Wise,
  ``Heavy quark physics,''
  Camb.\ Monogr.\ Part.\ Phys.\ Nucl.\ Phys.\ Cosmol.\  {\bf 10} (2000) 1.


\bibitem{CM}
N.~R.~Constable and R.~C.~Myers, JHEP {\bf 9911} (1999) 020
[arXiv:hep-th/9905081];


\bibitem{Gubser}
S.~S.~Gubser, [arXiv:hep-th/9902155].



\bibitem{add2} A.~Kehagias and K.~Sfetsos,
Phys.\ Lett.\ B {\bf 454} (1999) 270 [arXiv:hep-th/9902125];

\bibitem{Gubser2} V. Borokhov, S.S. Gubser,  JHEP {\bf
0305} (2003) 034, [arXiv: hep-th/0206098].

\bibitem{Kaidalov}
 A.~B.~Kaidalov,
``Hadronic Mass Relations From Topological Expansion And String Model,''
  Z.\ Phys.\ C {\bf 12} (1982) 63.


\bibitem{Simonov}
Y.~A.~Simonov,
 ``Heavy -- light mesons and quark constants f(B), f(D) in the method of vacuum
 correlators,''
  Z.\ Phys.\ C {\bf 53} (1992) 419.



\end{thebibliography}
\end{document}